\documentclass{blois}

\usepackage{fancyhdr,amsmath}
\usepackage{graphicx}
\usepackage{xspace}
\usepackage{rotating}
\usepackage{braket}
\def\Journal#1#2#3#4{{#1} {\bf #2}, #3 (#4)}

\def\PLB{{\em Phys. Lett.}  B}

\def\PRD{{\em Phys. Rev.} D}

\def\be{\begin{equation}}
\def\ee{\end{equation}}
\def\bea{\begin{eqnarray}}
\def\eea{\end{eqnarray}}

\def\mev{\,{\rm MeV}}
\def\gev{\,{\rm GeV}}

\newcommand{\Photo}{}

\begin{document}
  \begin{flushright}
    ADP--16--41/T997
  \end{flushright}
\vspace*{4cm}
\title{EW baryogenesis via DM}

\author{ Marek Lewicki }

\address{
ARC Centre  of  Excellence  for Particle  Physics  at  the  Terascale  (CoEPP) \& CSSM, Department of Physics, University of Adelaide, South Australia 5005, Australia\\
Institute of Theoretical Physics, Faculty of Physics, University of Warsaw, Pasteura 5, 02-093 Warsaw, Poland
}
\maketitle\abstracts{
We discuss the impact of a swifter cosmological expansion induced by modified cosmological history
of the universe on scenarios realising electroweak baryogenesis.
We detail the possible experimental bounds one can place on such cosmological modification and show how the detection capabilities of particle models are modified within these bounds.
On the particle physics side we focus on the Standard Model supplemented by a dimension
six operator which directly modifies the Higgs potential.
We show that due to the cosmological modification, electroweak baryogenesis in this model can be realized, with the modification of the triple Higgs coupling below HL-LHC sensitivity.}
\section{Introduction}

We will discuss electroweak baryogenesis \cite{Kuzmin:1985mm,Cohen:1993nk} in which the observed baryon asymmetry of the universe is created during the electroweak phase transition (EWPT). 
One of the requirements for this scenario is a strong first order phase transition which provides the departure from thermal equilibrium needed to create the asymmetry.
In the Standard Model with a $125$ GeV Higgs the transition is second order and an extension of the model is necessary\cite{Arnold:1992rz}. We will discuss a very simple and generic scenario where the SM Higgs potential is modified with a single nonrenormalisable operator  $\phi^6/\Lambda^2$ modifying the potential.  
 
The main issue we wish to discuss is the impact of non-standard cosmological history on models of EWBG.
We will again focus on a simple and generic modification assuming an additional component in the energy density of the Universe that red-shifts faster than radiation.
Such a modification of early cosmological history is rather poorly constrained by experiments and as we will show the new component can dominate the universe during the EWPT without contradicting any observations.  
 
Rather than discussing the production of the asymmetry during the phase transition we will discuss the so called sphaleron bound. This is a necessary condition for EWBG coming from the fact that the $SU(2)$ sphaleron processes which can create the asymmetry during the transition can also wash it away as the universe goes back to thermal equilibrium after the transition.
To avoid the wash-out the phase transition needs to be strongly enough first order to decouple the sphaleron processes. Usually this condition is satisfied in various models by modifying the potential to obtain a wide enough barrier between the symmetric minimum and the electroweak symmetry breaking one at the transition temperature. This also gives hope for detection of EWBG models as the high temperature Higgs potential is tightly bound to the zero temperature one whitch we can now probe at the LHC~\cite{Katz:2014bha}.
However decoupling of the sphalerons can also be aided by cosmological freeze-out if the expansion rate of the Universe during the transition is larger than in the usually assumed radiation domination case~\cite{Joyce:1997fc}.
Our main aim is to show how the detection capabilities of EWBG models change due to the modification of cosmological history~\cite{Lewicki:2016efe,Artymowski:2016tme}. 

\section{The Particle Model}\label{sec:particlemodel}
\begin{figure}[t] 
\begin{center}
\includegraphics[height=6cm]{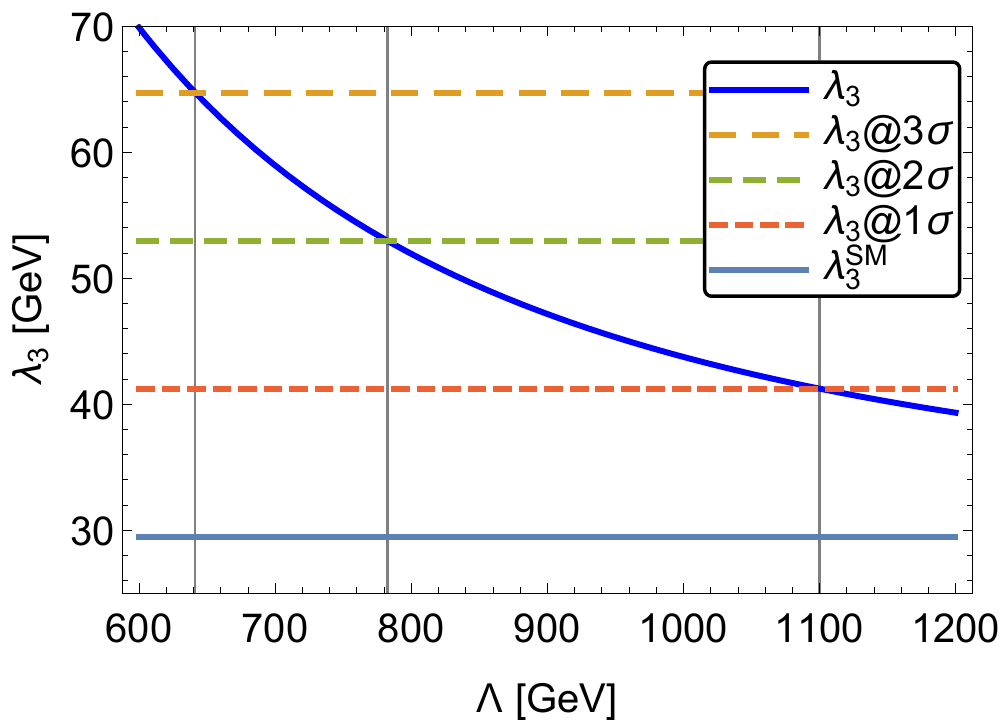} 
\end{center}
\caption{
The triple Higgs coupling $\lambda_3$ as a function of the cut-off scale (dark blue line), along with the HL-LHC sensitivity at $1$, $2$ and $3 \sigma$ (dashed lines). 
Thin vertical lines show cut-off scales corresponding to these sensitivities which are $\Lambda \approx 1102, 783$ and $641$ GeV, respectively.
\label{lambda3mplot}
}
\end{figure}

In this section we will describe the simple extension of the SM by a single nonrenormalisable operator $|H|^6$ suppressed by a certain mass scale $\Lambda$. We consider the potential
\begin{equation}\label{eqn:classpot}
V(H)=-m^2|H|^2+\lambda|H|^4+\frac{1}{\Lambda^2} |H|^6.
\end{equation}
The physical Higgs boson $\phi$ has the following tree level potential

\begin{equation}\label{eqn:classpottree}
V(\phi)^{\textrm tree}=-\frac{m^2}{2}\phi^2+\frac{\lambda}{4}\phi^4+\frac{1}{8}\frac{\phi^6}{ \Lambda^2}.
\end{equation}
The modified values of the SM parameters $\lambda$ and $m$ are set by the first and second derivative of the potential which are 
needed to correctly predict observed mass of the Higgs 
 and masses the gauge bosons given by the Higgs vev. 
%
%
The only affected measurable Higgs property is the triple-Higgs coupling given by
the third derivative of the effective potential.
This coupling can be measured at the LHC in double Higgs production events.
However, due to very low cross-section  high-luminosity experiments are required for a reliable measurement.
Upcoming high-luminosity phase of the LHC (HL-LHC) will be able to measure the value of $ \lambda_3 $ with roughly $ 40\% $ accuracy \cite{ATLAS-Collaboration:2012jwa}. Figure~\ref{lambda3mplot} shows the modification of $\lambda_3$ in our model, and the HL-LHC sensitivity at $1$, $2$ and $3 \sigma$. 

\section{Modification of the cosmological history and the sphaleron bound}

We will discuss a simple and generic modification of cosmological history that can effectively describe majority of existing cosmological models. 
We simply introduce a new energy constituent $\rho_N$ that red-shifts faster than radiation. 
That is the first Friedmann equation, reads
\begin{equation}\label{eq:friedmann}
H^2=\left( \frac{\dot{a}}{a} \right)^2 = \frac{8\pi}{3 M_{p}^2}\left( \frac{\rho_N}{a^n} + \frac{\rho_R}{a^4}\right),
\end{equation}
with $n > 4$.

There is one experimental constraint one can put on the expansion rate in the early universe which comes from Big Bang Nucleosynthesis.
At the temperature of $1$ MeV Neutrons have to freeze-out so that a precisely known fraction of them will remain to give us the observed abundances of light elements after recombination.
This gives us the expansion rate which is consistent with a universe filled with just SM radiation. However, within experimental uncertainty there is still room for a small fraction of an additional component such as our $\rho_N$. This result usually described in terms of effective number of neutrinos can be simply converted to a possible modification of the Hubble rate with respect to its value from the SM radiation $H/H_R|_{\rm BBN}\approx 1.018$~\cite{Lewicki:2016efe,Agashe:2014kda}.

\begin{figure}[t]
\begin{center}
\includegraphics[height=6.5cm]{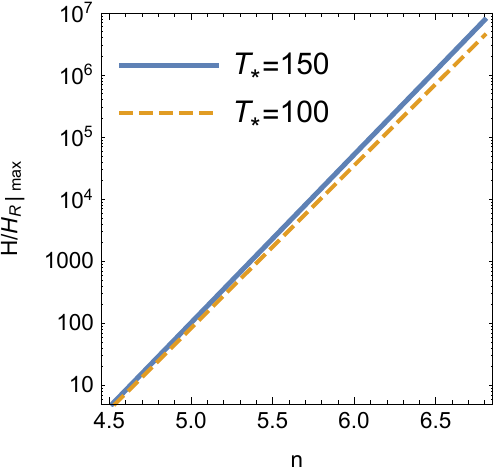} 
\includegraphics[height=6.5cm]{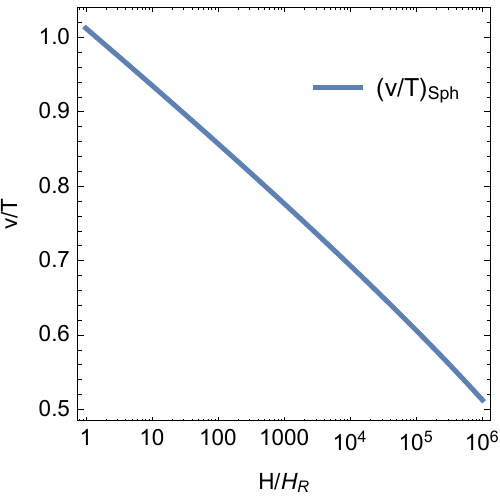} 
\end{center}
\caption{
Left panel: maximal experimentally allowed modification of the Hubble parameter for temperature values of interest.
 Right panel: the value of $v/T$ required to preserve the baryon asymmetry after the transition.
\label{HHRmaxandvoverT}
}
\end{figure}

The key point is that the new component contribution grows quickly as we go back in time and can easily dominate the expansion at earlier times. In order to find the expansion at the temperature of the phase transition $T_*$ we solve the simplified Friedmann equation neglecting radiation
\begin{equation}
\frac{H}{H_R} =
\sqrt{\left(\left. \frac{H}{H_R} \right|_{\rm BBN} \right)^2-1}
\left(\left(\frac{g_{*}}{g_{\rm BBN}}\right)^{\frac{1}{4}}\frac{T_*}{T_{\rm BBN}}\right)^{\frac{n-4}{2}},
\end{equation} 
where $T_ {\rm BBN}=1$\mev \ and the number of relativistic degrees of freedom is equal to $g_{\rm BBN}=10.75$ and $g_{*}=106.75$. Figure~\ref{HHRmaxandvoverT} shows the resulting maximal increase of the expansion rate in the temperature range close to the electroweak phase transition $T\in [100,150]\gev$.

In order to avoid washing away the asymmetry created during the phase transition, $SU(2)$ sphalerons have to be decoupled after the transition has commenced.
The simplest criterion for this decoupling simply requires the sphaleron rate to be smaller than the Hubble rate
\begin{equation}
\Gamma_{\rm Sph} = T^4 \mathcal{B}_0 \frac{g}{4 \pi} \left(\frac{v}{T}\right)^7 \exp \left( -\frac{4 \pi}{g}\frac{v}{T} \right)\leq H.
\end{equation}
Choosing $\mathcal{B}_0$ such that in the standard radiation dominated case this bound simply corresponds to $v/T\geq 1$ allows us to find $v/T$ required for decoupling as a function of $H/H_R$ which is shown in the right panel of Figure~\ref{HHRmaxandvoverT}.
\section{Results and conclusion}
\begin{figure}[t] 
\begin{center}
\includegraphics[height=6.5cm]{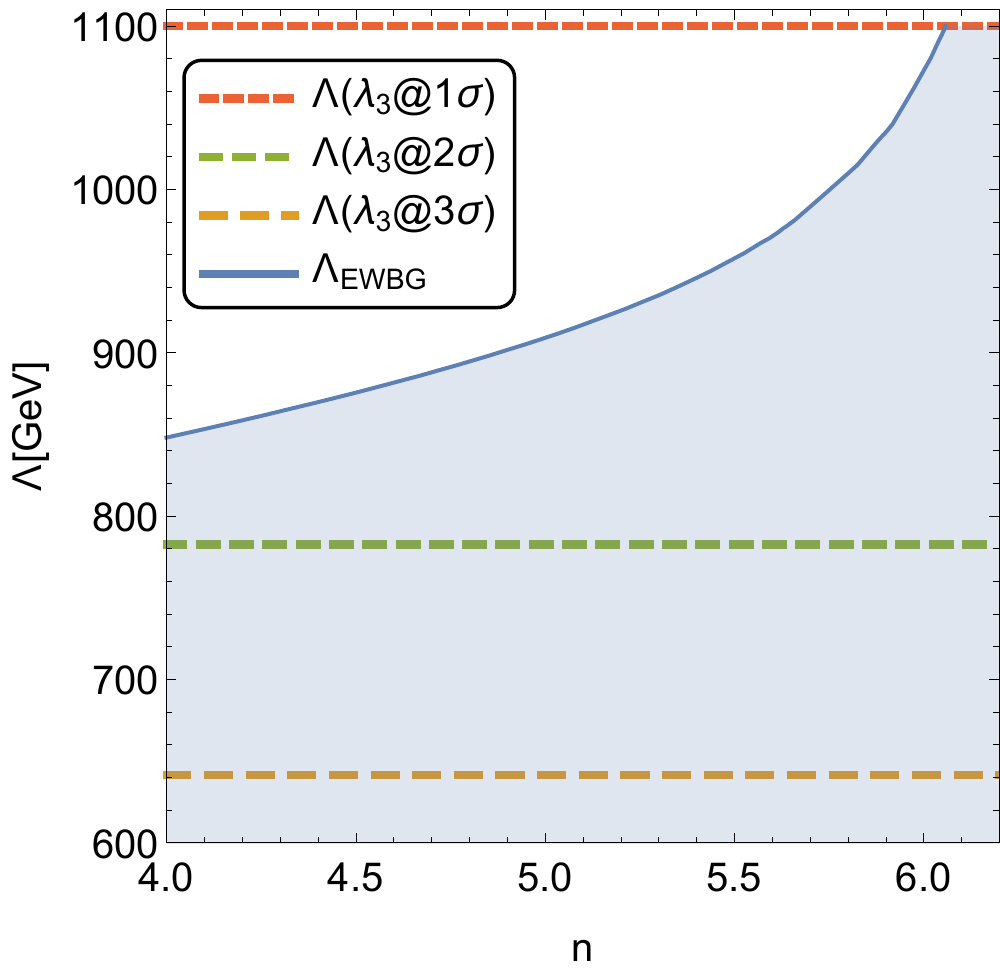} 
\includegraphics[height=6.5cm]{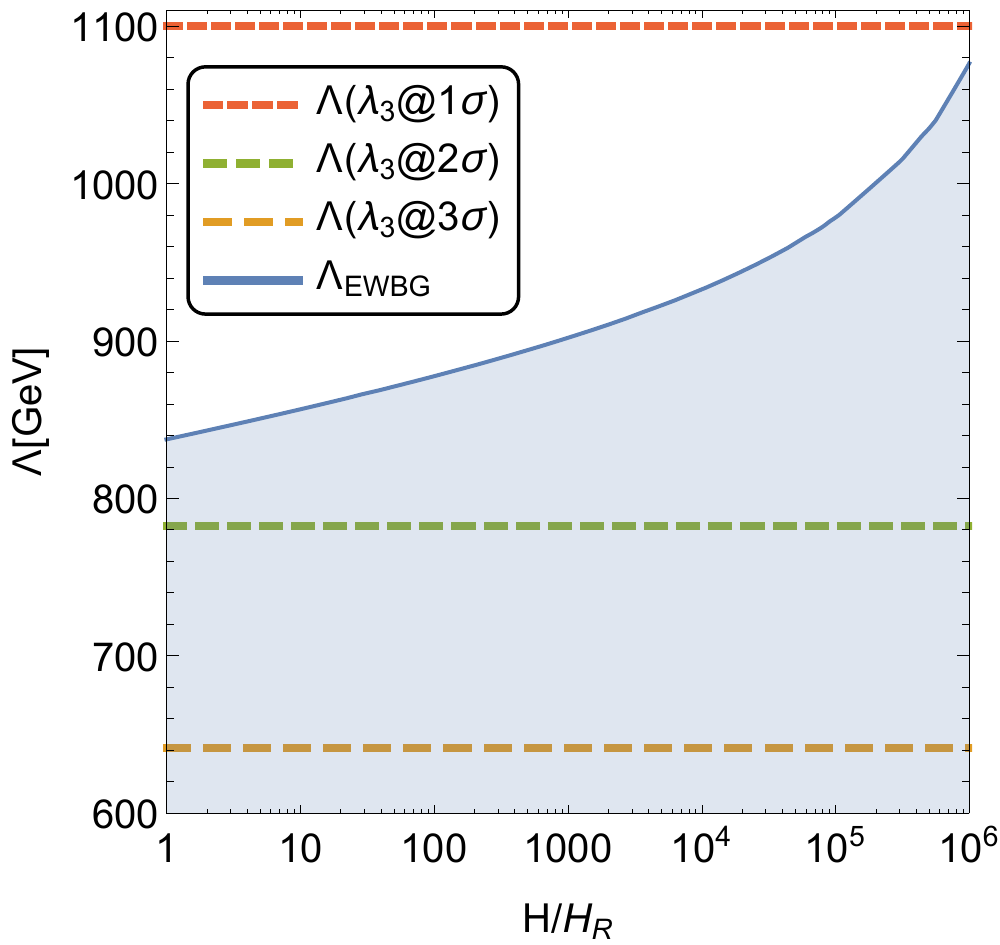} 
\end{center}
\caption{
Left panel: cut-off scales $\Lambda$ required for succesfull baryogenesis (blue region) as a function of $n$ which determines our cosmological model,
together with HL-LHC experimental reach (dashed horizontal lines).
Right panel: cut-off scales $\Lambda$ required for successful EWBG (blue region) for $n=6$ as a function of expansion rate.
\label{finalplot}
}
\end{figure}

We are finally in the position to use results from the previous sections and show how the new physics cut-off scale changes due to our cosmological modification. The values of the cut-off allowing bayrogenesis are shown in Figure~\ref{finalplot} together eperimental reach of the HL-LHC.  
 
The modification of the required new physics scale is numerically small, about 20\% for $\rho_N\propto a^{-6}$ leading to $\Lambda \approx 1100$ GeV.
However, for a cut-off larger than $1100$ GeV the phase transition would be of second order and thus we are actually circumventing the sphaleron bound altogether.
Our key result is that one only has to require a phase transition strong enough to create the asymmetry as its wash-out can always be avoided due to modification of cosmological history. 

It is also important to point out that for $n=6$ such a modification can be explained by a very light scalar field completely decoupled from the SM which at later time plays the role of dark matter~\cite{Lewicki:2016efe,Li:2013nal}.  

\section*{Acknowledgments}
I would like to thank James Wells, Tania Rindler-Daller and Micha\l \ Artymowski for fruitful collaboration.
This work was partially supported by the University of Adelaide, and the Australian Research Council through the ARC Centre of Excellence for Particle Physics at the Terascale (CoEPP) (CE110001104).
This work was partially supported by the Polish National Science Centre under research grant 2014/13/N/ST2/02712 and scholarship 2015/16/T/ST2/00527.

\end{document}